\documentstyle[aps,prl,epsfig]{revtex}
\begin{document}
\twocolumn[\hsize\textwidth\columnwidth\hsize\csname %
@twocolumnfalse\endcsname
 
\title{$c$-axis Optical Conductivity in Cuprates}
\author{A. Ram\v sak$^{1,2}$, I. Sega$^1$, and  P. Prelov\v sek$^{1,2}$,}
\address{ $^{1}$ J. Stefan Institute, 1000 Ljubljana, Slovenia }
\address{$^{2}$ Faculty of Mathematics and Physics, University of
Ljubljana, 1000 Ljubljana, Slovenia }
\date{\today}
\maketitle
\begin{abstract}\widetext
We investigate the $c$-axis optical conductivity and d.c. resistivity
of cuprate superconductors in the normal state. Assuming that the
interlayer hopping is incoherent we express the conductivity with
planar spectral functions obtained (i) from angle-resolved
photoemission experiments, (ii) using marginal Fermi liquid ansatz,
and (iii) with the finite-temperature Lanczos method for finite
two-dimensional systems described by the $t$-$J$ model. Here in the
low doping regime a pseudo-gap opening in the density of states
appears to be responsible for a semimetallic-like behavior of
$\rho_c(T)$.  In the optimally doped regime we find an anomalous
relaxation rate $\tau_c^{-1} \propto \omega + \xi_c T$. Analytically
this result is reproduced with the use of the marginal Fermi liquid
ansatz for the self energy with parameters obtained from the exact
diagonalization results.
\end{abstract}
 ]
\narrowtext

\section{Introduction}

The CuO$_2$ planes are common to all cuprate superconductors and
clearly determine most of the normal-state electronic
properties. The quantity which most
evidently displays the anisotropy of a particular material is the
optical conductivity $\sigma(\omega,T)$ and the corresponding
d.c. resistivity $\rho(T)$ \cite{coop}.

We present here the theoretical analysis of 
$\sigma_c(\omega)$ within the
standard planar $t$-$J$ model and we additionally allow for the 
hopping between layers,
\begin{equation}
H=\sum_l H_l^{(t{\rm -}J)} -t_c\sum_{l i
s}(c^\dagger_{lis}c_{l\!+\!1 is}+{\rm {H.c.}}).
\label{hop}
\end{equation}
where $H_l^{(t{\rm -}J)}$ is the $t$-$J$ model hamiltonian for the
layer $l$ and $t_c$ is hopping matrix element between layers $l$ and
$l\pm1$. The interlayer hopping matrix element $t_c$ is weak resulting
in an incoherent $c$-axis transport.

In the limit of weak interlayer hopping $t_c \ll t$ it is
straightforward to evaluate the dynamical $c$-axis conductivity
$\sigma_c(\omega)$ using the linear response theory. The result is
\cite{prelovsek98}
\begin{eqnarray}
\sigma_c(\omega)&=& {\sigma_c^0\over \omega}\int
{d\omega^\prime}
[f(\omega^\prime)-f(\omega^\prime+\omega)] \nonumber \\
&&\times {4 \pi t^2\over N}\sum_{{\bf k}} { A}({\bf k},\omega^\prime)
{A}({\bf k},\omega^\prime+\omega), \label{sigc}
\end{eqnarray}
where $\sigma_c^0= e_0^2 t_c^2 c_0/\hbar a_0^2 t^2$ is a
characteristic $c$-axis conductivity scale. Here $e_0$ is unit charge,
$a_0$ and $c_0$ are $ab$-lattice constant and $c$-axis
interlayer distance, respectively, $N$ is the number of sites
and $f(\omega)$ is the Fermi function.  This approximation 
assumes the independent electron propagation in each layer, where
electrons are described with the spectral function $A({\bf
k},\omega)$. Eq.~(\ref{sigc}) could be compared with a
related problem of interlayer hopping \cite{mott}, where the
corresponding expression for $\sigma_c^n(\omega)$ is obtained as a
convolution of the planar {\it density of states} (DOS) ${{ N_p}}(\omega)=
2/N \sum_{{\bf k}} { A}({\bf k},\omega-\mu)$ and $\mu$ is the
chemical potential,
\begin{eqnarray}
\sigma_c^n(\omega)&=&{\sigma_c^0\over \omega}
\int d\omega^\prime
[f(\omega^\prime)-f(\omega^\prime+\omega)]  \nonumber \\
&&\times{\pi t^2}{ {\cal N}}(\mu+\omega^\prime){{\cal
N}}(\mu+\omega^\prime+\omega), \label{sign}
\end{eqnarray}
appearing e.g. in disordered systems \cite{mott}, together with a
substitution for $t_c$ in the definition of $\sigma_c^0$ with some average
$\bar{t_c}$.  In cuprates both
alternatives have a justification, since the disorder introduced by
dopands residing between layers modifies the hopping elements. Hence
one can expect that the actual conductivity is a linear combination of
$\sigma_c$ and $\sigma_c^n$.

The knowledge of planar $A({\bf k},\omega)$ [or
${N_p}(\mu+\omega)$] should thus suffice for the evaluation of
$\sigma_c(\omega)$.  

\section{Results}

We test first the Mott formula~\cite{mott} for the d.c. resistivity
$\rho_c(T)=1/\sigma_c(\omega=0,T)$ by inserting experimentally
obtained photoemission spectra $\propto N_p(\omega)$ for
La$_{2-x}$Sr$_x$CuO$_4$ \cite{ino}. The density of states is
symmetrized around the Fermi energy and the result is presented in
Fig.~1(a) for various values $x=0.074 - 0.3$. Note that in the
underdoped regime semimetallic behavior is qualitatively consistent with
measured $\rho_c(T)$ \cite{coop}. In the optimally doped regime,
however, experiments for $\rho_c(T)$ give linear $T$-dependence. For
this regime therefore we take planar spectral functions consistent
with the marginal Fermi liquid (MFL) concept \cite{varm}, with
parameters for the self energy taken from exact diagonalization
\cite{jpspec}.  In Fig.~1(b) we present results for $\rho_c(T)$ 
from Eq.~(\ref{sigc}) for the
MFL scenario with full line and $\rho_c(T)$ from the Mott formula,
Eq.~(\ref{sign}), with dashed line.

Finally we present in Fig.~1(c) the d.c. resistivity $\rho_c(T)$ where
we use numerical results for $A({\bf k},\omega)$ and ${N_p}(\omega)$,
for systems with $N=16,18, 20$ sites \cite{jpspec}.  In the regime of
intermediate and even higher doping $c_h \geq 3/16$ $\rho_c(T)$ is
metallic-like for all $T$ while for the underdoped limit due to the
opening of the quasi-gap in the density of states $\rho_c(T)$ exhibits
a semimetallic-like behavior. 

To conclude, we point out that our approach gives a qualitative
agreement with experiments in LSCO, what gives support to such a
minimum model. Our results also confirm the general experimental
observation that $\rho_c(T)$ changes from a metallic one to a
semimetallic one, $d\rho_c(T)/dT < 0$, by decreasing doping.

\newpage
\begin{figure}[b]
\begin{center}\leavevmode  
\vskip -1.2  cm    
\includegraphics[width=0.74\linewidth,angle=-90]{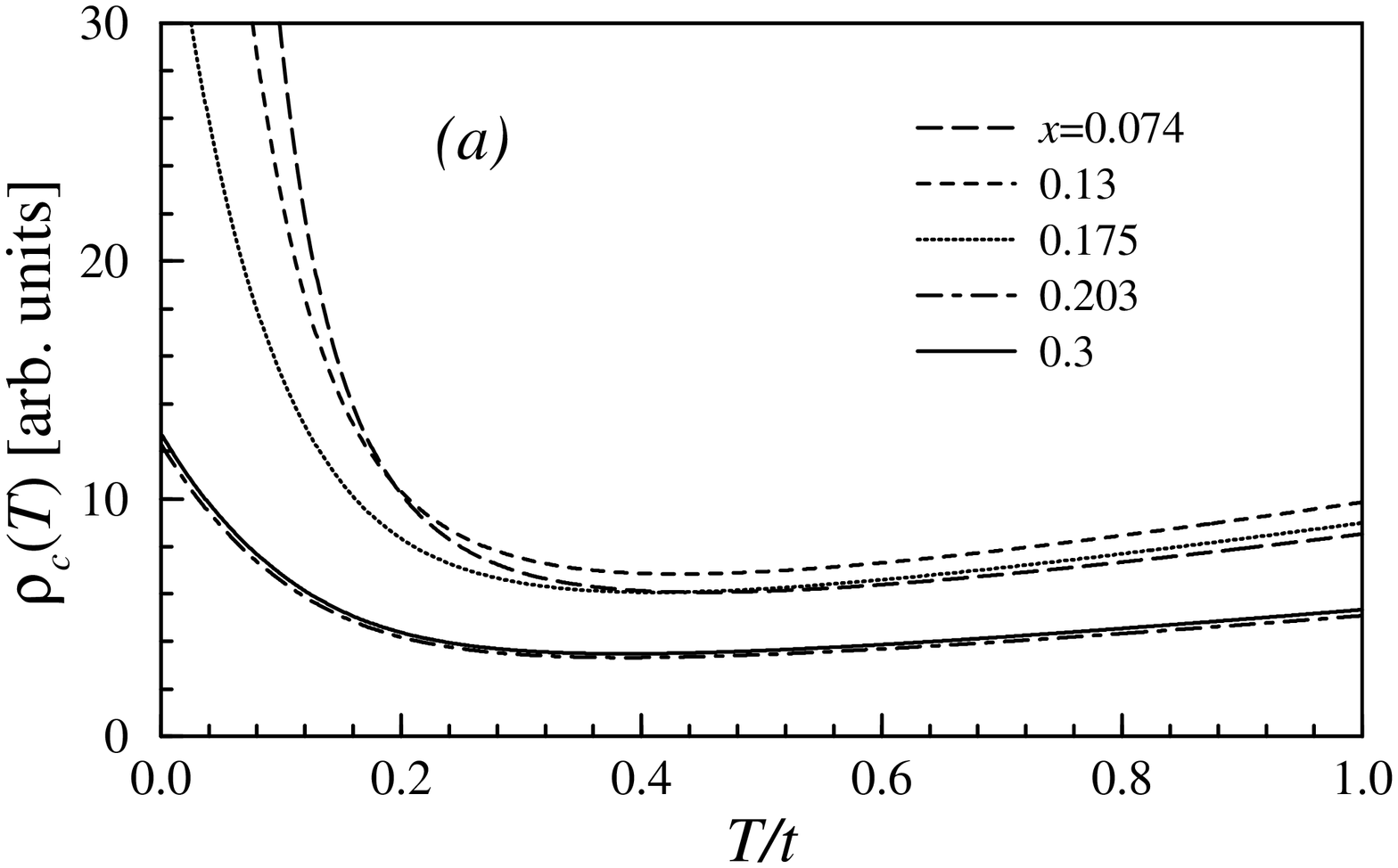}
\vskip -1.9  cm  
\includegraphics[width=0.74\linewidth,angle=-90]{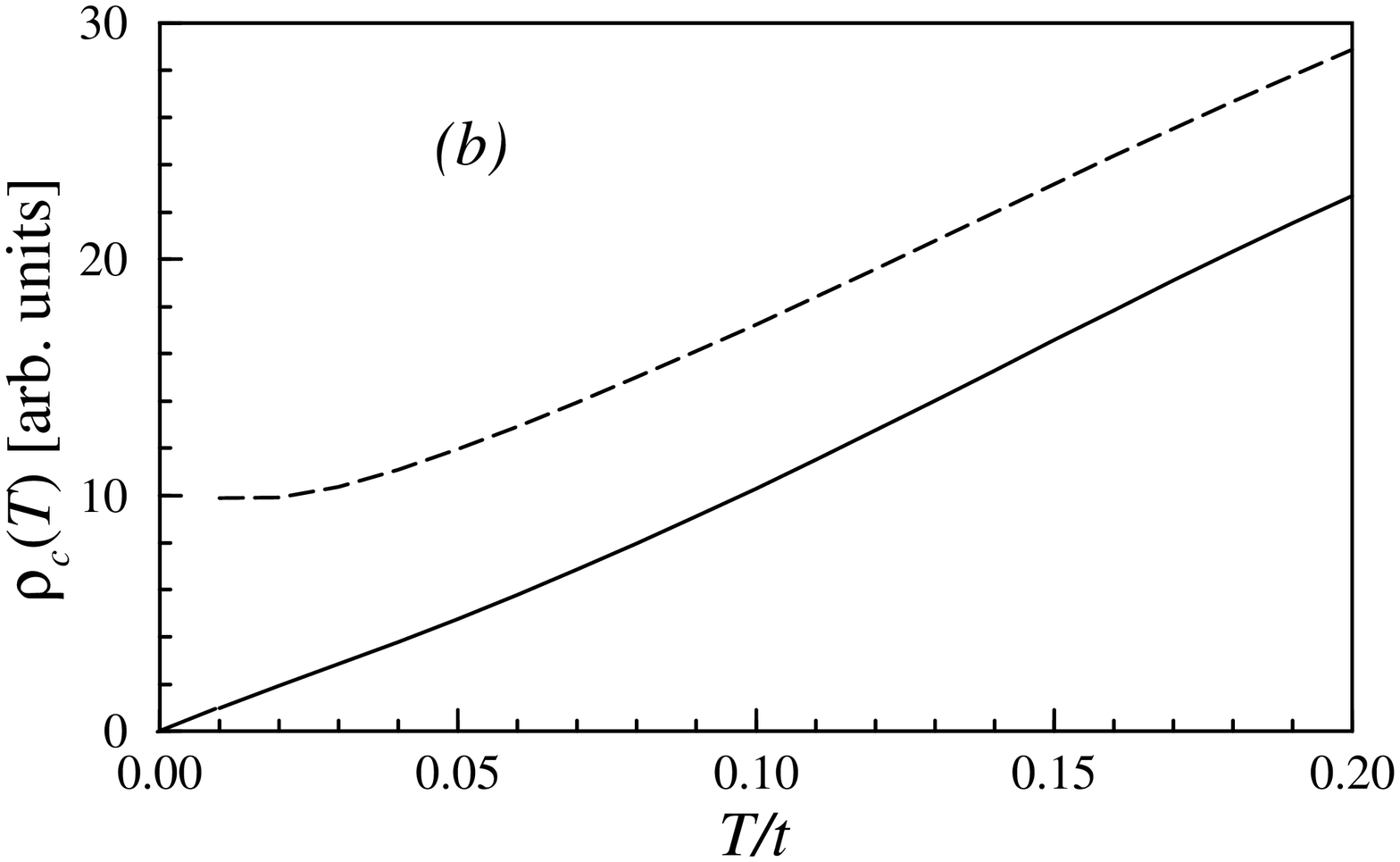}
\vskip -1.9  cm  
\includegraphics[width=0.74\linewidth,angle=-90]{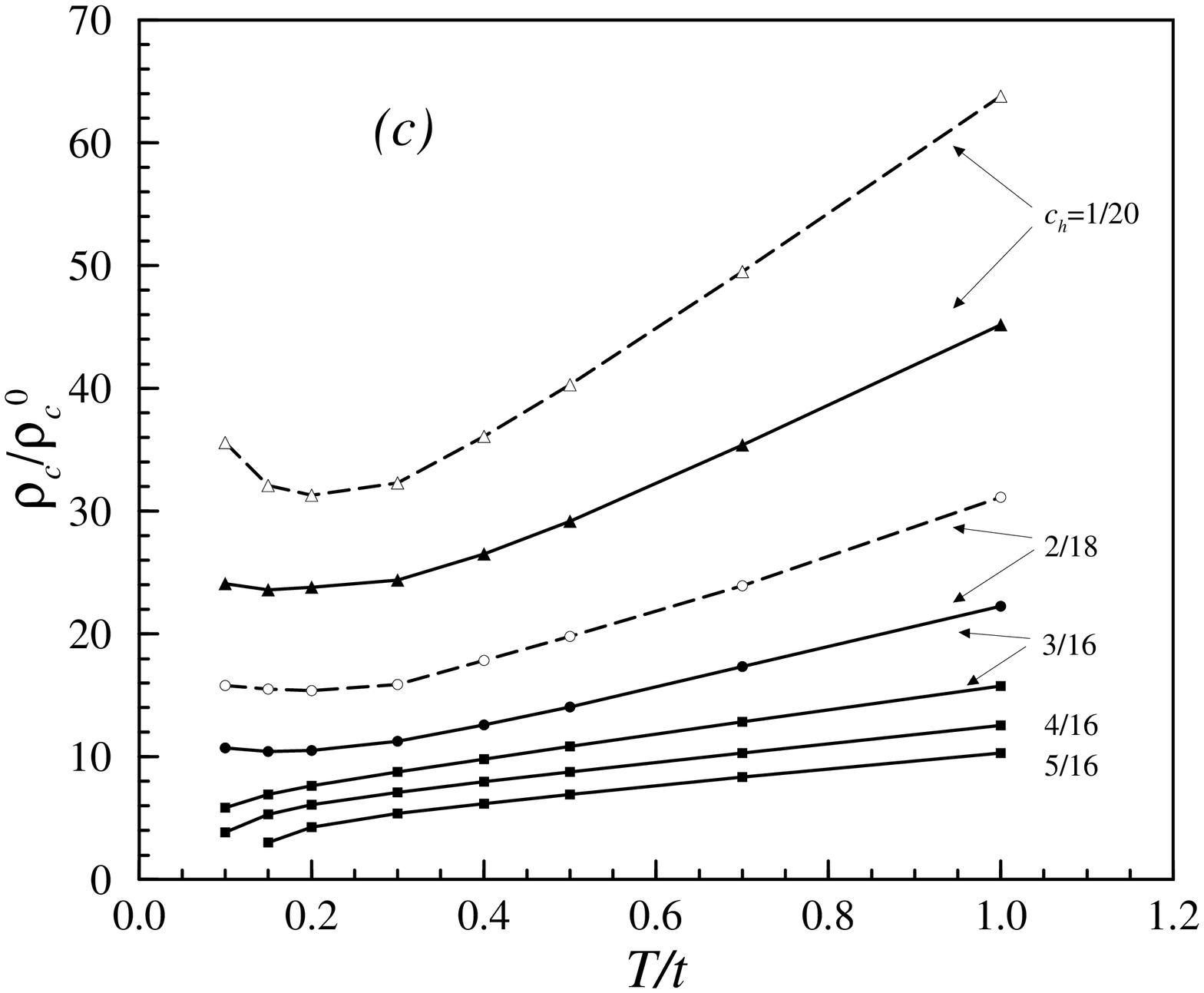}

\caption{ 
$\rho_c(T)$: 
(a) from ARPES $N_p$ for various doping $x$ 
(b) from MFL $A_{\bf k}(\omega)$ (full line) 
and from $N_p$ (dashed line)
(c) from $A_{\bf k}(\omega)$ (full lines) 
and $N_p$ (dashed lines) obtained 
with the $t$-$J$ model and various doping $c_h$.
}\label{figurename}\end{center}\end{figure}


\begin{references}
\bibitem{coop} For a review see e.g. S. L. Cooper and K. E. Gray, in
{\it Physical Properties of High Temperature Superconductors IV},
ed. by D. M. Ginsberg (World Scientific, Singapore, 1994), p.61.
\bibitem{prelovsek98} {P. Prelov\v sek, A. Ram\v sak, and I. Sega,
Phys. Rev. Lett. {\bf 81}, 3745 (1998).}
\bibitem{mott} N. F. Mott and E. A. Davis, {\it Electronic processes in
non-crystalline materials}, Clarendon Press, 1979.
\bibitem{jpspec} J. Jakli\v c and P. Prelov\v sek, Phys. Rev. B {\bf
49}, 5065 (1994); for a review see  J. Jakli\v c and P. Prelov\v sek,
cond-mat 9801333, to appear in Adv. Phys.;
J. Jakli\v c and P. Prelov\v sek, Phys. Rev. B {\bf
55}, R7307 (1996); P. Prelov\v sek, J. Jakli\v c, and K. Bedell,
Phys. Rev. B, to appear (1999).
\bibitem{ino} A. Ino {\it et al.}, Phys. Rev. Lett. {\bf 81}, 2124 (1998).
\bibitem{varm} C. M. Varma {\it et al.}, Phys. Rev. Lett. 
{\bf 63}, 1996 (1989).
\end{references}
\end{document}